\documentclass[12pt]{iopart}
\usepackage{graphicx}
\usepackage{citesort}
\begin{document}

\topical[Elementary building blocks on silicon surfaces]{Elementary
structural building blocks encountered in silicon surface
reconstructions}

\author{Corsin Battaglia $^1$, Katalin Ga\'{a}l-Nagy $^2$, Claude Monney $^1$, Cl\'ement Didiot $^1$, Eike Fabian Schwier $^1$, Michael Gunnar Garnier $^1$, Giovanni Onida $^2$ and Philipp Aebi $^1$}

\address{$^1$ Institut de Physique, Universit\'e de Neuch\^atel,
2000 Neuch\^atel, Switzerland}

\address{$^2$ Dipartimento di Fisica and
European Theoretical Spectroscopy Facility (ETSF), Universit\`a di
Milano, 20133 Milano, Italy}

\ead{corsin.battaglia@unine.ch}

\begin{abstract}
Driven by the reduction of dangling bonds and the minimization of
surface stress, reconstruction of silicon surfaces leads to a
striking diversity of outcomes. Despite this variety even very
elaborate structures are generally comprised of a small number of
structural building blocks. We here identify important elementary
building blocks and discuss their integration into the structural
models as well as their impact on the electronic structure of the
surface.
\end{abstract}


\section{Introduction}

Understanding the structural and electronic properties of silicon
surfaces at the atomic scale is of tremendous scientific and
technological importance. It has been known since 1958 that atoms at
the surface of a semiconductor assume a different structure than
that of the bulk \cite{Farnsworth58}. The creation of a surface
results in broken chemical bonds, so called dangling bonds, pointing
towards the vacuum. Dangling bonds are energetically unfavorable
causing surface atoms to rearrange or reconstruct in order to lower
the total energy of the surface, which may result in highly complex
atomic architectures. The determination of the atomic structure
requires the complementary role of different experimental and
theoretical techniques and remains a formidable challenge. It took
26 years of combined effort to solve the atomic structure of the
famous Si(111)-(7$\times$7) reconstruction \cite{Takayanagi85}.

\begin{figure}
\centering
\includegraphics{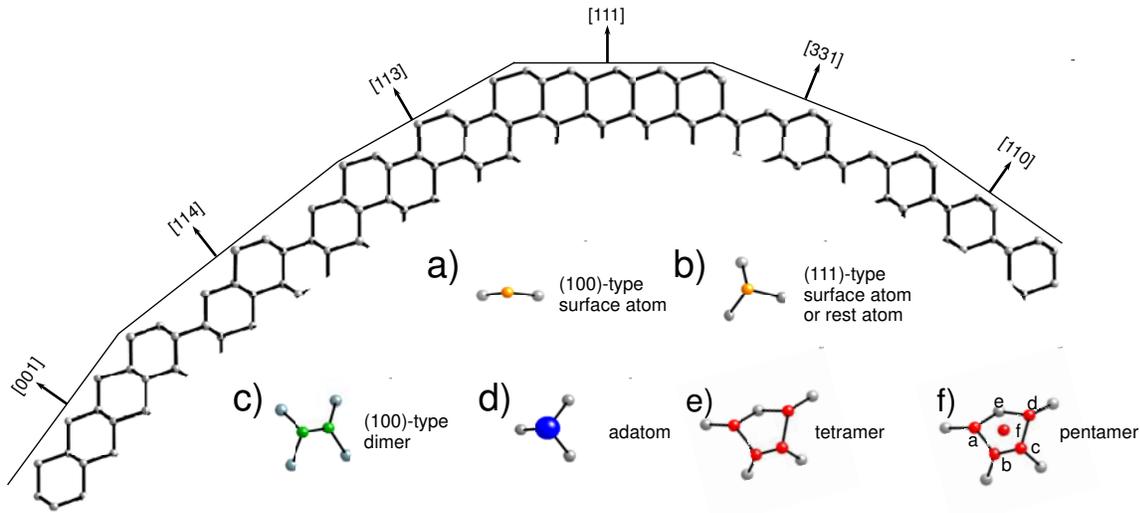}
  \caption{\label{fig:Overview} Side view of silicon crystal along the $[\bar{1}10]$ direction with various surfaces indicated. An overview on the elementary structural building blocks encountered in silicon surface reconstructions is also shown.}
\end{figure}

Surprisingly only a handful planar silicon surfaces with a stable
reconstruction is known \cite{Gai01b}. Most studies have
concentrated on surfaces with a surface normal between the [100] and
[110] direction including the (111) surface (see Fig.
\ref{fig:Overview} for an overview). Only little is known about
surfaces with orientations away from this plane.

Despite the high structural complexity of silicon surface
reconstructions one often encounters common elementary structural
building blocks (Fig. \ref{fig:Overview}(a)-(f)). Identifying these
building blocks is important not only for a better understanding of
these surfaces but may also serve as a guide for the elaboration of
new structural models. In this short review we describe the most
important silicon surface reconstructions emphasizing the role of
these elementary structural building blocks. First we discuss the
properties of the bulk-truncated surfaces and identify two
prototypical types of surface atoms. Then we focus on the strategies
adopted by the various surfaces in order to reduce the number of
dangling bonds by integrating these building blocks into more
complex structures. We also include a discussion on the relation
between their structural and electronic properties.



\section{Bulk-truncated surfaces}

In bulk silicon, each of the tetrahedrally coordinated atoms forms
four covalent bonds with its four nearest neighbors. Each bond
contains two paired electrons. When a surface is formed, some of
these bonds will be broken, leading to unsaturated orbitals, the
so-called dangling bonds, containing only one unpaired electron. The
lack of electron pairing makes dangling bonds unstable. The atoms in
the surface region will move away from their bulk positions trying
to minimize the surface energy. When this happens, the surface is
said to relax or reconstruct depending on how the surface atoms seek
new coordinates. Surface relaxation refers to the case when surface
atoms are displaced from their bulk positions, but there is no
change in the surface periodicity. Surface reconstruction on the
other hand refers to atomic displacements causing the symmetry
parallel to the surface to be lower than that of the bulk.

At metal surfaces, the electrons are free to rearrange their
distribution in space. Relaxation by adjustment of the interlayer
spacing of the first few atomic planes is often sufficient to
minimize the surface energy. At semiconductor surfaces, the truly
directional chemical bonds between atoms lead to considerable
elastic strain which increases the total energy of the surface.
Stable surface reconstructions are obtained when the strain energy
is compensated by the energy gain which results from the reduction
of dangling bonds.

When studying the energetics of semiconductor surface
reconstructions, it is important to take into account the electronic
structure of the surface as well. Further lowering of the total
energy may be achieved when surface states are either empty or fully
occupied by two electrons. Spontaneous symmetry breaking and the
related lifting of degeneracies, opening a gap between electronic
states, is often at the origin of the driving force for surface
reconstruction. In fact, besides a few exceptions, the bandstructure
of reconstructed semiconductor surfaces tends to be semiconducting.
In subsequent sections, several mechanism leading to a gap in the
electronic spectrum of the surface will be discussed.

 A thorough understanding of the geometry of the
non-reconstructed bulk-truncated surfaces is the starting point for
the elaboration of any structural model. Fig. \ref{fig:Si100}(a),
\ref{fig:Si111}(a), \ref{fig:Si114}(a), \ref{fig:Si113}(a),
\ref{fig:Si110}(a), and \ref{fig:Si331}(a) present the dangling bond
configuration of the bulk-truncated Si(100), Si(111), Si(114),
Si(113), Si(110), and Si(331) surfaces respectively, whose
reconstructions will be discussed in the following. Yellow colored
atoms represent surface atoms with dangling bonds. Two prototypical
types of surface atoms may be distinguished. The (100)-type surface
atoms found on the (100) bulk-truncated surface (Fig.
\ref{fig:Si100}(a)) and the (111)-type surface atoms occurring on
the (111) bulk-truncated surface (Fig. \ref{fig:Si111}(a)).
(100)-type surface atoms carry two dangling bonds and have two
backbonds to the substrate, while the (111)-type surface atoms carry
only one dangling bonds and share three backbonds. These two types
of surface atoms also occur on surfaces with other orientations. On
bulk-truncated Si(114) (Fig. \ref{fig:Si114}(a)) and Si(113) (Fig.
\ref{fig:Si113}(a)) surfaces, both types of surface atoms exist
simultaneously, whereas on the Si(110) (Fig. \ref{fig:Si110}(a)) and
on the Si(331) (Fig. \ref{fig:Si331}(a)) surface only (111)-type
surface atoms occur. In the following we discuss several strategies
with which the silicon surfaces reduce the number of dangling bonds.

\section{Strategy 1: Dimers}

\begin{figure}
\centering
\includegraphics{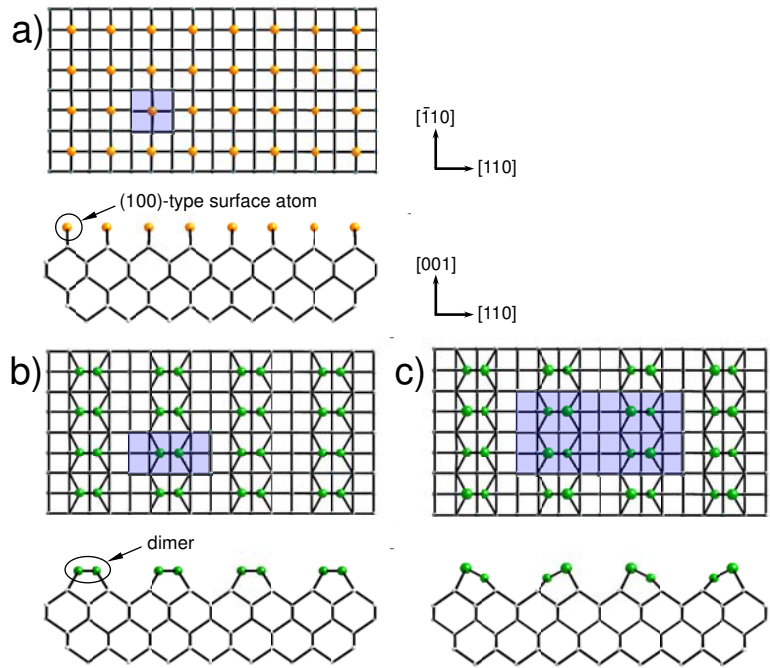}
  \caption{\label{fig:Si100}(a) Bulk-truncated Si(100)-(1$\times$1) surface. (b) Si(100)-(2$\times$1) with symmetric dimers. (c) Si(100)-c(4$\times$2) with asymmetric dimers. The corresponding unit cells are shown in blue. }
\end{figure}

A conceptually simple strategy to reduce the number of dangling
bonds is the formation of dimers found on the Si(100) surface. In
1958 Farnsworth \textit{et al.} \cite{Farnsworth58} reported that
low-energy electron diffraction (LEED) of clean Si(100) produces
half-integral diffraction spots indicating a (2$\times$1)
periodicity in real space. However, it was not before 1992, after
the publication of the first low-temperature scanning tunneling
microscopy (STM) images \cite{Wolkow92} that a general consensus
about its detailed atomic structure emerged \cite{Dabrowski00}. In
order to minimize their energy, surface atoms on clean Si(100) move
pairwise towards each other and form a new bond resulting in
symmetric dimers shown in Fig. \ref{fig:Si100}(b). However,
low-temperature STM images \cite{Wolkow92,Hata02} clearly showed
that the dimers on Si(100) are buckled, i.e. the two atoms of the
dimer have a different height above the surface plane (see Fig.
\ref{fig:Si100}(c)). These asymmetric dimers are also supported by
LEED \cite{Over97} and other techniques \cite{Dabrowski00}.
Asymmetric dimers prefer higher periodicities, (2$\times$2) and
c(4$\times$2), which appear because the direction of buckling of
neighboring dimers is correlated. Unless stabilized by surface
defects, correlation is partially destroyed around 200 K. Above this
temperature LEED usually sees an average (2$\times$1) order. At room
temperature, dimers appear symmetric in STM images, since thermal
vibrations flip the buckling direction of dimers faster than what
can be observed by STM.

How does the formation of dimers influences the electronic structure
of the surface? When two surface atoms pair up to form a dimer only
one of the two dangling bonds carried by each (100)-type surface
atom gets eliminated. The orbitals, associated with the electrons
participating in the formation of the dimer bond, overlap resulting
in a bonding $\sigma$ and antibonding $\sigma^*$ combination
\cite{Dabrowski00}. Since the overlap is large, the energy splitting
between the two states is large, causing the occupied $\sigma$ state
and the empty $\sigma^*$ state to become broad resonances in the
valence and the conduction band respectively. The remaining two
dangling bonds mix into a $\pi$ and $\pi^*$ bond \cite{Dabrowski00}.
For symmetric dimers the energy splitting between these two states
is small resulting in a partial overlap, which renders the system
metallic. The formation of asymmetric dimers allows a slight energy
gain and opens a gap between the $\pi$ and $\pi^*$ state, which
renders the system semiconducting in agreement with experiment.

\section{Strategy 2: Adatoms and rest atoms}

\begin{figure}
\centering
\includegraphics{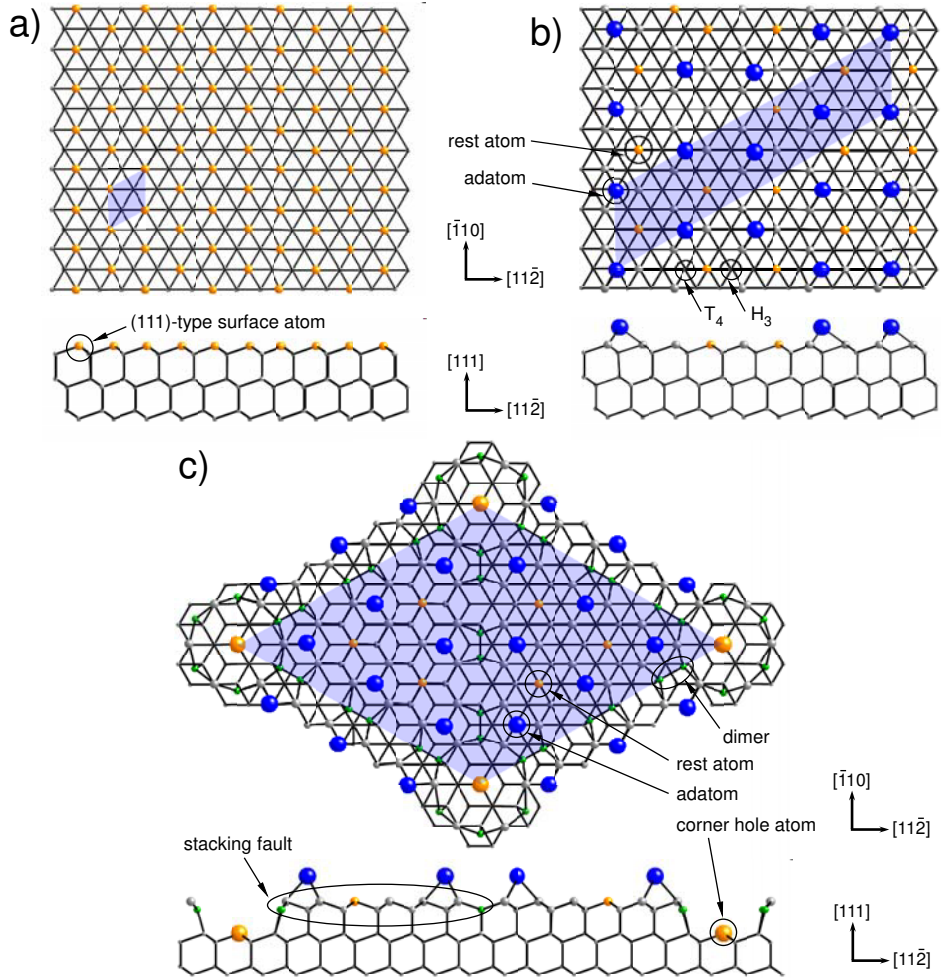}
  \caption{\label{fig:Si111}(a) Bulk-truncated Si(111)-(1$\times$1) or Ge(111)-(1$\times$1) surface. (b) Ge(111)-c(2$\times$8) reconstruction. (c) Si(111)-(7$\times$7) reconstruction.}
\end{figure}

Prototypical adatoms in combination with rest atoms are encountered
on (111) surfaces of the elemental semiconductors. Here we first
discuss the Ge(111)-c(2$\times$8) reconstruction (Fig.
\ref{fig:Si111}(b)), because it is less complex than the famous
Si(111)-(7$\times$7) reconstruction (Fig. \ref{fig:Si111}(c)). The
Ge(111)-c(2$\times$8) reconstruction was identified in 1963
\cite{Lander63,Jona65,Palmberg67} by LEED. On the bulk-truncated
surface (111)-type surface atoms are arranged in a hexagonal pattern
and are only second-nearest neighbors (Fig. \ref{fig:Si111}(a)).
Their nearest neighbors are three subsurface atoms, with which they
share a bond. In order to reduce the number of dangling bonds on the
surface, additional germanium atoms, called adatoms (blue atoms in
Fig. \ref{fig:Si111}(b)), saturate each three adjacent dangling
bonds by forming three bonds, called backbonds, with the three
nearest surface atoms \cite{Chadi81}. Its fourth orbital carrying a
single electron points towards the vacuum. An adatom thus replaces
three dangling bonds by a new dangling bond.

Adatoms may occupy two possible sites indicated in Fig.
\ref{fig:Si111}(b). These geometries are distinguished as hollow
($H_3$) and atop ($T_4$) sites depending on whether the substrate
atom below the adatom is found in the fourth or second layer. In
$H_3$ sites the adatom is three-fold coordinated, in $T_4$ sites the
adatom is approximately four-fold coordinated due to the substrate
atom directly below in the second layer. The unambiguous
discrimination between adatoms in $T_4$ and $H_3$ sites was finally
achieved by x-ray diffraction in 1990 \cite{vanSilfhout90} favoring
$T_4$ sites.

Although each adatom reduces the number of dangling bonds, it is not
favorable to saturate a surface with the maximum number of adatoms.
Of the 16 dangling bonds per c(2$\times$8) unit cell of the
bulk-truncated Ge(111) surface, four adatoms saturate 12. Each
adatom still carries one remaining dangling bond. So the number of
dangling bonds per reconstructed c(2$\times$8) unit cell is 8. The
four surface atoms (yellow atoms in Fig. \ref{fig:Si111}(b)), whose
dangling bonds have not been saturated by adatoms are called rest
atoms. The structure is further stabilized by an electronic charge
transfer from the adatoms to the rest atoms
\cite{Meade89,Takeuchi92,Takeuchi95} resulting in fully filled rest
atom states and empty adatom states in agreement with experiment
\cite{Becker89,Hirschorn91}.

The balance between lowering in energy due to the reduction of
dangling bonds and the energy increase caused by the bond distortion
is very delicate \cite{Meade89,Bechstedt01}. Compared to the
Ge(111)-c(2$\times$8) reconstruction, the Si(111)-(7$\times$7)
reconstruction is much more complex. Since its discovery in 1959
\cite{Schlier59} using LEED the (7$\times$7) reconstruction has
become the prototype for studying complex reconstructions occurring
at semiconductor surfaces.

A mystery for many years, the atomic structure of
Si(111)-(7$\times$7) has been resolved by Takayanagi \textit{et al.}
\cite{Takayanagi85,Takayanagi85b} in 1985 on the basis of
transmission electron diffraction data, assisted in part by the
observation of adatoms in STM images by Binnig \textit{et al.}
\cite{Binnig83}. Their now widely accepted dimer-adatom-stacking
fault (DAS) model shown in Fig. \ref{fig:Si111}(c) consists of 12
silicon adatoms in the first layer (blue atoms), a stacking fault
bilayer (second and third layer), within which 9 dimers (green
atoms) in the third layer border the triangular faulted and
unfaulted subunits. Note that the dimers observed on
Si(111)-(7$\times$7) are not the same as the ones observed on
Si(100)-(2$\times$1). Whereas the two atoms of a standard (100)-type
dimer carry each one remaining dangling bond, the (111)-type dimers
are completely saturated. A deep vacancy, called the corner hole is
located at each apex of the unit cell (on top of the large yellow
atom). The 6 three-fold bonded atoms in the second layer falling in
between the adatoms of each triangular subunit are rest atoms (small
yellow atoms).

The DAS model reduces the number of dangling bonds from 49 for the
unreconstructed (7$\times$7) unit cell to 19 (12 dangling bonds for
the adatoms, 6 dangling bonds for the rest atoms and one dangling
for the atom below the corner hole). These 19 dangling bonds deliver
14 electrons which fill the energetically lower lying rest atoms and
corner hole states, i.e. 7 electrons are transferred from the adatom
states to the rest atom and corner hole states. The remaining 5
electrons remain in the adatom bands resulting in a metallic
semiconductor surface in agreement with experiment
\cite{Hamers86,Losio00}.

\begin{figure}
\centering
\includegraphics{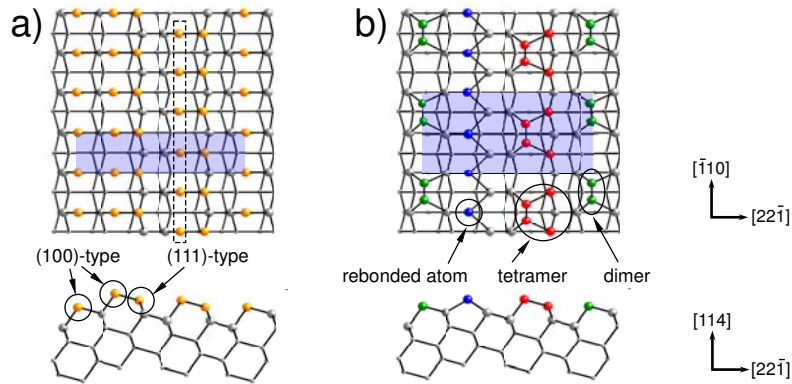}
  \caption{\label{fig:Si114}(a) Bulk-truncated Si(114)-(1$\times$1) surface. (b) Si(114)-(2$\times$1) reconstruction.}
\end{figure}

\section{Strategy 3: Tetramers and pentamers}

Tetramers and pentamers are more complex schemes to eliminate
dangling bonds. Tetramers are found on Si(114) and Si(113). We first
discuss the conceptually simpler Si(114)-(2$\times$1)
reconstruction.

The Si(114)-(2$\times$1) reconstruction was first reported in 1993
\cite{Suzuki93}. A structural model was proposed by Erwin \textit{et
al.} in 1996 \cite{Erwin96}. On the bulk-truncated Si(114) surface
shown in Fig. \ref{fig:Si114}(a) both (111)-like and (100)-like
surface atoms are observed carrying one and two dangling bonds
respectively. The (100)-like surface atoms marked by a dashed
rectangle in Fig. \ref{fig:Si114}(a) dimerize. Due to the immediate
vicinity of two neighboring (111)-like surface atoms bonding to the
dimer atoms, the dimer is topologically different from the standard
dimer encountered on the Si(100) surface. The resulting structure
containing the dimer atoms plus the two (111)-like surface atoms is
called a tetramer (red atoms in Fig. \ref{fig:Si114}(b)). An
additional standard dimer is formed by pairs of (100)-like surface
atoms (green atoms). The remaining four surface atoms per unit cell
are replaced by two so-called rebonded atoms (shown in blue), each
having three backbonds and one dangling bond.

The reconstructed Si(114) surface unit cell exhibits a total of 8
dangling bonds, 4 on the tetramer, 2 on the two rebonded atoms and 2
on the standard (100)-like dimer. The 8 electrons coming from these
8 dangling bonds fill four surface states. Two unoccupied surface
states separated by a small gap from the filled part of the spectrum
have also been identified in a first-principles study
\cite{Smardon04}.

\begin{figure}
\centering
\includegraphics{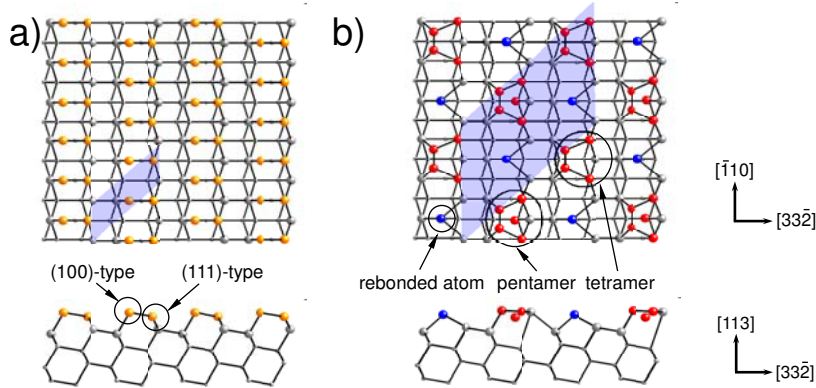}
  \caption{\label{fig:Si113}(a)  Bulk-truncated Si(113)-(1$\times$1) surface. (b) Si(113)-(3$\times$2) reconstruction.}
\end{figure}

We now turn to the Si(113) surface reconstruction reported in 1985
by Gibson \textit{et al.} \cite{Gibson85,footnote1}. The (113)
bulk-truncated surface consists of alternating rows of (111)- and
(100)-like surface atoms. The model proposed by Dabrowski \textit{et
al.} in 1994 \cite{Dabrowski94} is shown in Fig. \ref{fig:Si113}(b).
It shares several building blocks with the Si(114)-(2$\times$1)
reconstruction, namely tetramers and rebonded atoms. Tetramers and
rebonded atoms are arranged in a (3$\times$1) order such that they
alternate along $[\bar{1}10]$. Taking into account the dangling bond
on the rebonded atom and the 4 dangling bonds on the tetramer, we
end up with 5 dangling bonds per unit cell. Thus from an
electon-counting point of view this structure is expected to be
metallic in contrast to the semiconducting nature of the surface
observed in experiments. To solve this problem, every second
tetramer captures an interstitial silicon atom (atom $f$ in Fig.
\ref{fig:Overview}(f)) leading to the (3$\times$2) periodicity
observed at room temperature \cite{Dabrowski00,Dabrowski94}. This
interstitial atom affects the electronic structure of Si(113),
changing the metallic (3$\times$1) into a semiconducting
(3$\times$2) surface, however inducing considerable strain. The
vertical component of the strain is relieved as the structure
relaxes strongly towards the vacuum, elevating a pentagonal ring of
atoms. This pentamer, formed by the original tetramer and the common
neighbor of the two (111)-like surface atoms, is almost flat and
parallel to the surface. Above 800 K, interstitial atoms hop from
one tetramer to the other \cite{Laracuente98}. Due to the missing
correlation between hopping interstitial atoms, the periodicity
observed by LEED is (3$\times$1).

\begin{figure}
\centering
\includegraphics{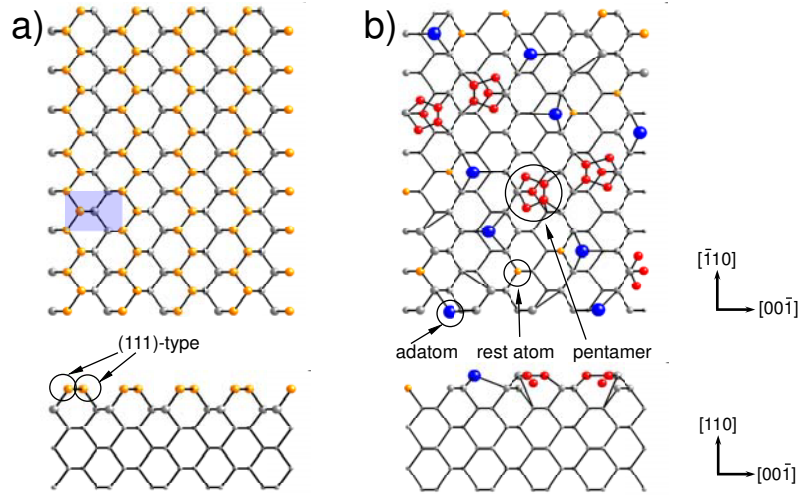}
  \caption{\label{fig:Si110}(a) Bulk-truncated Si(110)-(1$\times$1) surface. (b) Si(110)-(16$\times$2) reconstruction.}
\end{figure}

A structural model containing a slightly different pentamer has been
proposed by An \cite{An00} to explain the pentagons observed in STM
images of the (16$\times$2) reconstruction of Si(110)
\cite{footnote2}. Already in 1965 Jona \cite{Jona65} studied the
Si(110) surface by LEED and observed a structure with a periodicity
of "possibly 16". The bulk-truncated Si(110) surface consists of
double rows of (111)-like surface atoms running along the
$[\bar{1}10]$ direction. Since all surface atoms of the
bulk-truncated Si(110) surface are (111)-like, the original pentamer
found on the Si(113) surface must be modified. In the
adatom-tetramer-interstitial (ATI) model
\cite{An00,Stekolnikov04b,Stekolnikov04} shown partially in Fig.
\ref{fig:Si110}(b) (for the complete structural model including the
steps see Ref. \cite{Stekolnikov04b}), pentamers are formed by four
adatoms (atoms $a,b,c$, and $d$ in Fig. \ref{fig:Overview}(f))
forming the tetramer, one surface atom ($e$), and the interstitial
atom ($f$). In contrast to the pentamers encountered on the Si(113)
surface, only atom $e$, which serves as an anchor point for the
pentamer, is provided by the surface atoms. All other atoms
($a,b,c,d$, and $f$) are additional atoms which must be provided by
the step which is an integral part of the unit cell. Besides the
pentamers, adatoms accompanied by rest atoms are found to interlink
the double rows of surface atoms in a complex way. Although we
encounter rest atoms on surfaces away from the [111] direction, it
is important to note that rest atoms are always (111)-type surface
atoms carrying only one dangling bond. Since (100)-type surface
atoms carry two dangling bonds, which render them energetically
highly unstable, they do not qualify as rest atoms and have not been
observed.

Using STM, we have recently been able to resolve for the first time
pentagons on the Si(331) surface \cite{Battaglia08}, very similar to
the ones observed on the Si(110) surface \cite{An00}. Since the
discovery of the Si(331)-(12$\times$1) reconstruction more than 17
years ago \cite{Wei91,footnote3}, several structural models
containing dimers and adatoms have been proposed
\cite{Olshanetsky98,Gai01}. However, none of these models is able to
explain the pentagons observed in our STM images.

\begin{figure}
\centering
\includegraphics{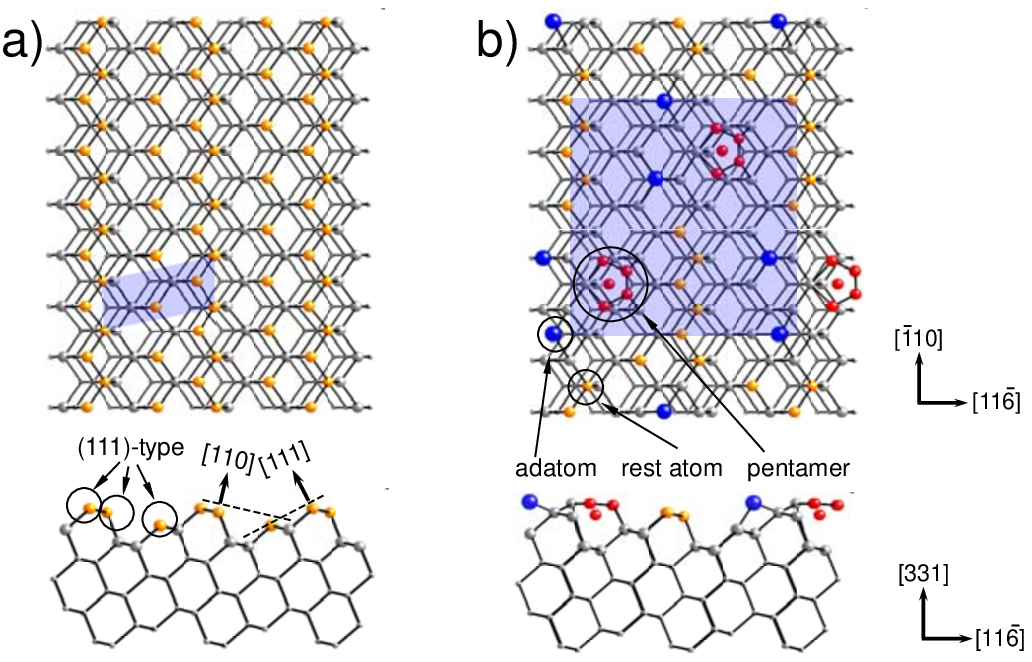}
  \caption{\label{fig:Si331}(a)  Bulk-truncated Si(331)-(1$\times$1) surface. (b) Si(331)-(12$\times$1) reconstruction.}
\end{figure}

Inspired by the structural model of the Si(110)-(16$\times$2)
reconstruction, we have proposed a new structural model for
Si(331)-(12$\times$1) containing silicon pentamers as essential
structural building blocks (see Fig. \ref{fig:Si331}b). The
arrangement of dangling bonds on the bulk-truncated surface of
Si(110) and Si(331) differ. Whereas (111)-type surface atoms on the
Si(110) surface occur in double rows running along the
$[\bar{1}10]$, double rows of (111)-type surface atoms alternate
with single rows on the bulk-truncated Si(331) surface. The
bulk-truncated Si(331) surface can actually be viewed as consisting
of small alternating (110) and (111) terraces (see Fig.
\ref{fig:Si331}(a)). In analogy with the Si(110) model the two
pentamers for the Si(331) model are anchored on the double rows of
surface atoms in a way such that the local bonding geometry is
exactly the same as on the (110) surface. Some of the remaining
dangling bonds are saturated by adatoms. Stekolnikov \textit{et al.}
\cite{Stekolnikov04b} have already noted for the Si(110) case that
it is energetically more favorable to leave some rest atoms
unsaturated than to introduce the maximum number of adatoms into the
model. This allows further reduction of the surface energy by
electron transfer from the adatom to the rest atom in analogy with
Ge(111)-c(2$\times$8) and Si(111)-(7$\times$7).

\section{Summary and conclusion}

Important lessons may be learned by analyzing and comparing existing
structural models for silicon surface reconstructions. Although each
surface adopts its own strategy to reduce the number of dangling
bonds, we identified elementary structural building blocks including
dimers, adatoms, rest atoms, rebonded atoms, tetramers, and
pentamers common to several reconstructions. We discussed their
integration into the structural models and the consequences on the
electronic structure.

\ack Stimulating discussions with Christophe Ballif, Steven C.
Erwin, Pascal Ruffieux and Wolf-Dieter Schneider are gratefully
acknowledged. This work was supported by the Fonds National Suisse
pour la Recherche Scientifique through Division II, the Swiss
National Center of Competence in Research MaNEP, and by the EU's 6th
Framework Programme through the NANOQUANTA Network of Excellence
(NMP-4-CT-2004-500198)

\section*{References}
\bibliography{Si331_BuildingBlocks}

\begin{thebibliography}{10}
\bibitem{footnote2}
The periodicity of the conventionally labeled Si(110)-(16$\times$2)
  reconstruction is more correctly described in matrix notation by {\small
  $\left( \begin{array}{cc} 11 & -5 \\ 2 & 2 \\ \end{array} \right)$}.
\bibitem{footnote3}
The periodicity of the conventionnally labeled Si(331)-(12$\times$1)
  reconstruction is more correctly described in matrix notation by
  {\small$\left( \begin{array}{cc} 2 & 0 \\ 1 & 6 \\ \end{array} \right)$}.
\end{thebibliography}

\end{document}